\documentclass[prl,a4paper,superscriptaddress,twocolumn,amsmath,amssymb]{revtex4-1}

\usepackage{graphicx}
\graphicspath{{./Figs/}}
\usepackage{color}
\usepackage{bm}
\usepackage{dsfont}
\usepackage{epstopdf}
\usepackage{hyperref}

\newcommand {\Fig}[1] {Fig.~\ref{#1}}

\newcommand{\phosphorus}{$^{31}$P}

\newcommand{\sitwonine}{$^{29}$Si}
\newcommand{\sitwoeight}{$^{28}$Si}

\newcommand{\tone}{$T_1$}

\newcommand{\tonee}{$T_{\rm{1e}}$}
\newcommand{\ttwon}{$T_{\rm{2n}}$}
\newcommand{\ttwoe}{$T_{\rm{2e}}$}

\newcommand{\natsi}{$^{\rm nat}$Si}

\begin{document}

\title{\sitwonine\ nuclear spins as a resource for donor spin qubits in silicon}

\author{Gary Wolfowicz}
\email{gary.wolfowicz@materials.ox.ac.uk}
\affiliation{London Centre for Nanotechnology, University College London, London WC1H 0AH, UK}
\affiliation{Dept.\ of Materials, Oxford University, Oxford OX1 3PH, UK}  

\author{Pierre-Andr\'e Mortemousque}
\affiliation{School of Fundamental Science and Technology, Keio University,	3-14-1 Hiyoshi, Kohoku-ku, Yokohama 223-8522, Japan}

\author{Roland Guichard}
\affiliation{Department of Physics and Astronomy, University College London, London WC1E 6BT, UK}

\author{Stephanie Simmons}
\affiliation{Centre for Quantum Computation and Communication Technology, The University of New South Wales, Sydney, NSW 2052, Australia}
\author{Mike L. W. Thewalt}
\affiliation{Dept.\ of Physics, Simon Fraser University, Burnaby, British Columbia V5A 1S6, Canada}

\author{Kohei M. Itoh}
\affiliation{School of Fundamental Science and Technology, Keio University,	3-14-1 Hiyoshi, Kohoku-ku, Yokohama 223-8522, Japan}

\author{John~J.~L.~Morton}
\email{jjl.morton@ucl.ac.uk}
\affiliation{London Centre for Nanotechnology, University College London, London WC1H 0AH, UK} 
\affiliation{Dept.\ of Electronic \& Electrical Engineering, University College London, London WC1E 7JE, UK} 



\begin{abstract}
Nuclear spin registers in the vicinity of electron spins in solid state systems offer a powerful resource to address the challenge of scalability in quantum architectures. We investigate here the properties of \sitwonine\ nuclear spins surrounding donor atoms in silicon, and consider the use of such spins, combined with the donor nuclear spin, as a quantum register coupled to the donor electron spin. We find the coherence of the nearby \sitwonine\ nuclear spins is effectively \emph{protected} by the presence of the donor electron spin, leading to coherence times in the second timescale --- over two orders of magnitude greater than the coherence times in bulk silicon. We theoretically investigate the use of such a register for quantum error correction, including methods to protect nuclear spins from the ionisation/neutralisation of the donor, which is necessary for the re-initialisation of the ancillae qubits. This provides a route for multi-round quantum error correction using donors in silicon. 
\end{abstract}

\maketitle

Modular `quantum network' architectures consisting of multiple quantum registers connected by interaction channels have emerged as a flexible, robust and scalable model for quantum computation. Such models typically assume high-fidelity operations which can be performed locally within the quantum registers (in contrast to potentially lossy channels between them), allowing operations such as local quantum error correction (QEC)~\cite{Moussa2011,Waldherr2014,Taminiau2014}, entanglement purification~\cite{Nickerson2013}, and even enhanced quantum sensing~\cite{Schaffry2011a,Ajoy2015}. This approach is well suited to spins of defects in the solid state, such as vacancies in diamond~\cite{Dutt2007} or silicon carbide~\cite{Widmann2014}, rare-earth dopants in various crystals~\cite{Wolfowicz2015} and donors in silicon~\cite{Morton2008}. Each of these offers a (sparse) environment of nuclear spins, in the vicinity of the defect spin, possessing potentially long coherence times. This has been explored recently using nitrogen-vacancies in diamond, first through the control of remote $\rm ^{13}C$ nuclear spins~\cite{Kolkowitz2012,Zhao2012,Taminiau2012} and later realizing a single round of quantum error correction (QEC)~\cite{Waldherr2014, Taminiau2014}. 

Naturally occurring silicon (\natsi) has three stable isotopes: $\rm ^{28}Si$ (92.2~\%), $\rm ^{29}Si$ (4.7~\%) and $\rm ^{28}Si$ (3.1~\%), where only \sitwonine\ has a non-zero spin ($I=1/2$) and could form part of a quantum register. In silicon, much recent focus has been on isotopically enriched \sitwoeight\ to \emph{remove} the \sitwonine\ spins \cite{Itoh2014}, leading to donor electron spin coherence times up to 3 seconds~\cite{Wolfowicz2013} and donor nuclear spin coherence times from minutes to hours~\cite{Saeedi2013, Muhonen2014}. The disadvantage of such \sitwoeight\ material is that the only additional resource for the donor electron spin is the nuclear spin of the donor itself.

Our focus here is on \natsi, and in particular the \sitwonine\ nuclear spins around the donor. Nuclear spin coherence times of \sitwonine\ have been studied in the absence of the donor electron (i.e.\ in bulk NMR~\cite{Dementyev2003}, or using a single \sitwonine\ atom coupled to a nano-device~\cite{Pla2014}) --- in such cases the nuclear spins can freely flip-flop and the Hahn echo \ttwon\ is limited to around 5~ms. However, the presence of the donor electron spin is known to form a `frozen core' of nuclear spins around the donor, changing the bath dynamics by detuning nuclear spins from their neighbours as a result of the spatially varying hyperfine coupling. For these reasons, one could expect the \ttwon\ of \sitwonine\ in the vicinity of the donor to be significantly longer --- an indication of this is in the \ttwon\ of the donor nuclear spin itself (strongly detuned from any of the neighbouring \sitwonine) which was reported to be about 1 second in \natsi\ \cite{Petersen2013}.

In this Letter, we consider the potential of both the donor nuclear spin and local \sitwonine\ spins as a register of qubits in silicon, characterising their coherence times and examining their use for local QEC. For QEC we consider both single-donor approaches (based on single-donor spin measurement devices~\cite{Pla2012,Pla2013,Pla2014}) and donor ensemble approaches (which could form part of hybrid architectures with superconducting resonators and qubits~\cite{Kubo2012, Wesenberg2009}). In addition to long coherence times, requirements for multi-round QEC include qubit manipulation and in particular the re-initialisation of ancilla qubits. Initialisation schemes (e.g.\ by single-spin measurement or optical hyperpolarisation~\cite{Steger2011,Pla2012}) involve the ionisation of the donor, and thus we conclude by examining how to ensure a nuclear spin data qubit can be made robust to this process.

We used a float-zone \natsi\ sample doped with \phosphorus\  at a concentration of $6 \times 10^{15}$~cm$^{-3}$. Measurements were performed at 4.5~K to obtain an electron \tone\ ($> 5$~s) sufficiently long compared to all other experimental timescales. Pulsed electron spin resonance (ESR) and
electron-nuclear double resonance (ENDOR) experiments were realised using a Bruker X-band Elexsys system ($\approx$~0.3 T, 9.7~GHz). The magnetic field was set parallel to the [001] Si crystal axis, where the electron spin coherence time \ttwoe\ is maximized ($\approx 0.5$~ms \cite{Tyryshkin2006}). Dynamical decoupling sequences applied to the nuclear spins were synthesised directly from an arbitrary waveform generator (Agilent 81180).


\begin{figure}[t]%
	\includegraphics[width=\columnwidth]{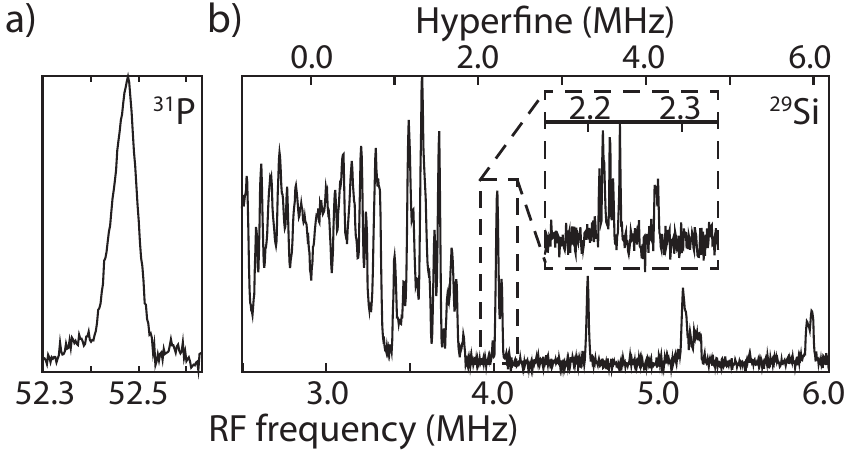}%
	\caption{\textbf{ENDOR spectra of \phosphorus\ and \sitwonine\ nuclear spins in \natsi\ at 344.2~mT.}
		(a)~Davies ENDOR spectrum of the \phosphorus\ donor in silicon. The peak at 52.475~MHz corresponds to a hyperfine of 117.53~MHz, and the linewidth of 60~kHz is consistent with previous ENDOR measurements in \natsi~\cite{Sekiguchi2014}. RF $\pi$ pulse length = 13$~\mu$s. (b) Davies ENDOR spectrum of \sitwonine. The hyperfine interaction values with the donor electron spin are calculated as twice the shift from the nuclear Zeeman frequency of 2.91~MHz, and ranging up to 6~MHz. Inset shows a high-resolution spectrum centred around $A/2= 2$~MHz, showing sub-components of the peaks due to the anisotropy of the hyperfine interaction. RF $\pi$ pulse length = 50$~\mu$s in main panel and 1.6~ms in inset.}
	\label{fig:29Sispectrum}%
\end{figure}

We begin by characterising the \phosphorus\ and \sitwonine\ nuclear spins through Davies ENDOR spectroscopy~\cite{Davies1974, Tyryshkin2006}. 
The \phosphorus\ donor nuclear spin has a well-known gyromagnetic ratio of 17.23~MHz/T and a hyperfine interaction with the donor electron spin of 117.53~MHz~\cite{Feher1959a}. \sitwonine\ spins in the bath around the donor have a gyromagnetic ratio of 8.46~MHz/T and hyperfine couplings to the donor electron spin of up to 6~MHz. An in-depth study of all couplings and related sites can be found in~\cite{Hale1969}. Spectral overlapping makes weakly coupled \sitwonine\ more difficult to distinguish experimentally --- for these, the hyperfine interactions can be simulated using the Kohn-Luttinger model of the electron wavefunction (see Supplementary Material).


\begin{figure}[t]%
	\includegraphics[width=\columnwidth]{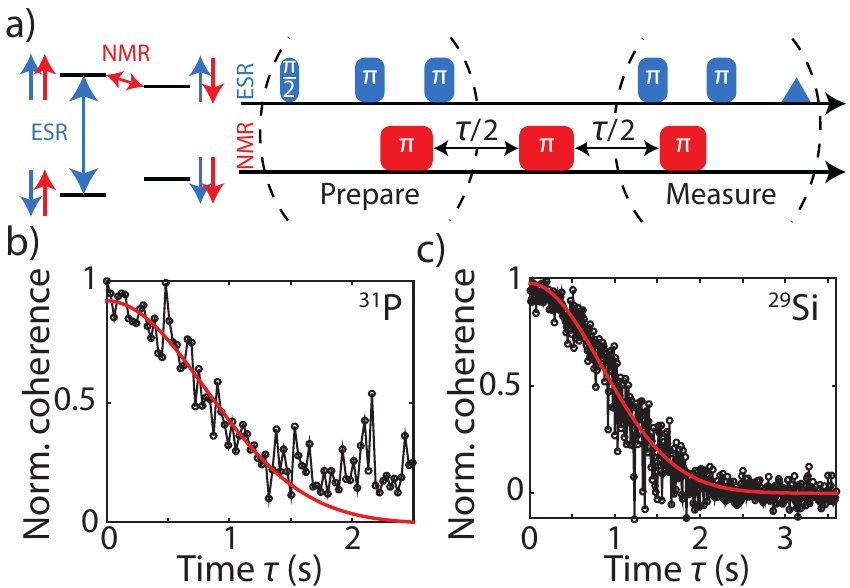}%
	\caption{\textbf{Nuclear spin coherence times of \phosphorus\ and \sitwonine.}
		(a)~Left: energy diagram for the donor electron spin coupled to a spin-1/2 nuclear spin. Right: nuclear spin coherence measurement sequence taken from Ref \cite{Morton2008}. The blocks defined by the dashed brackets move together when $\tau$ is varied.
		(b)~\phosphorus\ nuclear spin coherence decay. The signal shown is the magnitude of the ESR in-phase and quadrature detection, hence the fit (red) is constrained to decay to zero as the noise is always positive.
		(c)~Coherence decay for a \sitwonine\ nuclear spin with $A = 4.03$~MHz.
	}
	\label{fig:NuclearT2}%
\end{figure}

We then measure the coherence times (\ttwon) for these various nuclear spins, based on the approach of coherent state transfer from the donor electron spins to the nuclear spin, and back again at some later time, as shown in \Fig{fig:NuclearT2}(a) \cite{Morton2008}. Microwave pulses on the ESR transitions must be selective on a particular nuclear spin state, and thus have a bandwidth significantly less than the relevant hyperfine coupling. This is trivial in the case of the donor nuclear spin, however for \sitwonine\ spins this required microwave pulse lengths of $0.5~\mu$s.
Figure \ref{fig:NuclearT2}(b) shows the nuclear spin coherence decay is observed for \phosphorus\ with a resulting decay time $T_{\rm 2n} = 1.1 \pm 0.1$~s. A comparable coherence time of $1.22 \pm 0.03$~s was measured for a \sitwonine\ nuclear spin with hyperfine coupling $A = 4.03$~MHz (\Fig{fig:NuclearT2}(c)), notably over 200 times longer than in bulk natural silicon. In both cases, the decay followed a stretched exponential function $\exp \left( -\tau/T_{\rm 2n}\right)^n$ with stretch factor $n$ around 2, typical of decoherence from spectral diffusion in \natsi~\cite{Abe2010}.

\begin{figure}[t]%
	\includegraphics[width=\columnwidth]{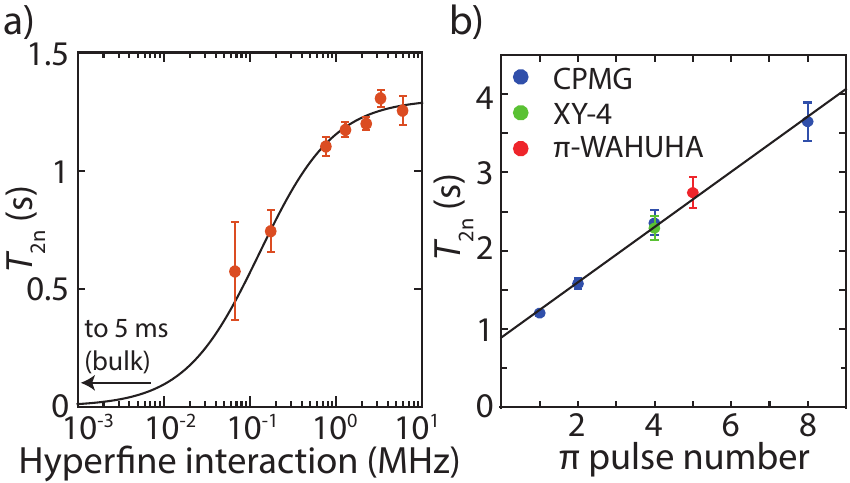}%
	\caption{\textbf{\sitwonine\ coherence time as a function of hyperfine coupling and dynamical decoupling (DD)}
		(a)~The \sitwonine\ coherence times, \ttwon, vary with the strength of the hyperfine coupling to the donor electron (and thus, indirectly, as a function of the distance between the two).
		The line in black is only a guide to the eye, limited at low coupling to the bulk NMR value (5~ms) and at large coupling to $\approx 1.3$~s.
		(b)~Measured \ttwon\ times under various DD sequences for a specific \sitwonine\ site with coupling $A = 2.23$~MHz. CPMG, $\pi$-WAHUHA, and XY-4 all offer identical protection of the nuclear spin coherence (in proportion to the number of refocusing pulses), showing that indirect flip-flops in the environment of the measured \sitwonine\ spin are responsible for decoherence. 
	}
	\label{fig:NuclearT2vsJ}%
\end{figure}

The \sitwonine\ nuclear spin coherence time was found to depend strongly on the hyperfine coupling to the donor electron spin, as shown in \Fig{fig:NuclearT2vsJ}(a). For the strongest hyperfine coupling $(A\gtrsim3~$MHz), the coherence time saturates at $\approx 1.3$~s, and then decreases with weaker coupling, towards the bulk NMR value of 5~ms~\cite{Dementyev2003}. 
Two decoherence mechanisms can be considered in this case. The first is due to flip-flop of \sitwonine\ spin pairs very far from the donor, and thus far from the measured \sitwonine\ spin. The distant \sitwonine\ spin pairs have negligible hyperfine interaction with the donor electron spin (compared to the dipolar coupling within the pair), and are therefore not detuned from one another, allowing for flip-flops \cite{Hayashi2008}. Their small coupling with the measured nuclear spin is compensated by the very large number of pairs involved in the process ($\approx 10^8$~\cite{Guichard2014}). A second process, recently proposed~\cite{Guichard2014}, involves only a few spin pairs that are much closer to the donor and are located at lattice sites that are equivalent by symmetry. Such pairs have equal coupling to the electron spin, hence there is no detuning within the pair. Both of these mechanisms induce decoherence to the measured nuclear spin via the Ising ($ZZ$) interaction, also termed an \emph{indirect} flip-flop process as it is due to flip-flops of neighbours
\footnote{In principle, decoherence could also proceed through flip-flops between the measured \sitwonine\ spins and their own equivalent pairs, however the number of such equivalent sites is low and this would be a weak process. Indeed, in \Fig{fig:NuclearT2vsJ}(a) the four most strongly coupled \sitwonine\ are from different lattice sites and have between 4 and 8 equivalent sites each, while their \ttwon\ vary insignificantly.}. We go on to use dynamical decoupling (DD) to further investigate these decoherence mechanisms.

Dynamical decoupling has been used extensively in different contexts ranging from extending coherence times \cite{Muhonen2014, Saeedi2013} and performing spectroscopy \cite{Bylander2011, Taminiau2012} to probing quantum interactions \cite{Zhao2011a, Ma2014}. In \Fig{fig:NuclearT2vsJ}(b), \sitwonine\ nuclear spins at a specific site ($A=2.23$~MHz) are subject to different types of DD sequence: CPMG~[Ref~\citep{Meiboom1958}], XY-4~[Ref~\citep{Gullion1990}] and a modified version of WAHUHA~[Ref~\citep{Haeberlen1968}]. CPMG consists of a train of $\pi$ pulses that refocuses $ZZ$ interactions between spins. Our experiments show that  under CPMG \ttwon\ increases linearly with the number of $\pi$ pulses, up to $3.7 \pm 0.2$~s (for eight $\pi$ pulses). This improvement provides an additional evidence that indirect flip-flops are the likely source of decoherence. This can be further tested using the WAHUHA sequence: 
$Y_{\pi} \frac{\tau}{} X_{\frac{\pi}{2}} \frac{\tau}{} X_{\pi} \frac{\tau}{}(-Y)_{\frac{\pi}{2}} \frac{2\tau}{}Y_{\pi} \frac{2\tau}{} Y_{\frac{\pi}{2}} \frac{\tau}{} X_{\pi} \frac{\tau}{} (-X)_{\frac{\pi}{2}} \frac{\tau}{} Y_{\pi}$, 
modified here (called $\pi$-WAHUHA) to also include $\pi$ pulses to allow for refocusing of inhomogeneous broadening ($T_2^*$). By alternating the rotation axis of the $\frac{\pi}{2}$ pulses, this refocuses the dipolar interaction between the measured nuclear spin and any equivalent pair. By comparison with the results from CPMG, it can be seen that this sequence does not improve the nuclear spin coherence beyond what would be expected from its five $\pi$ pulses, which eliminates the possibility of a decoherence mechanism due to direct flip-flops. Finally, XY-4, which has four $\pi$ pulses with alternating rotation axes, is applied to check for any effect from pulse errors, and unlike CPMG is a universal DD protocol required for use in general qubit applications. In summary, the coherence of both \phosphorus\ and \sitwonine\ nuclear spins have been measured to be in the order of seconds, and can be extended using DD sequences.

\begin{figure}[t]%
	\includegraphics[width=\columnwidth]{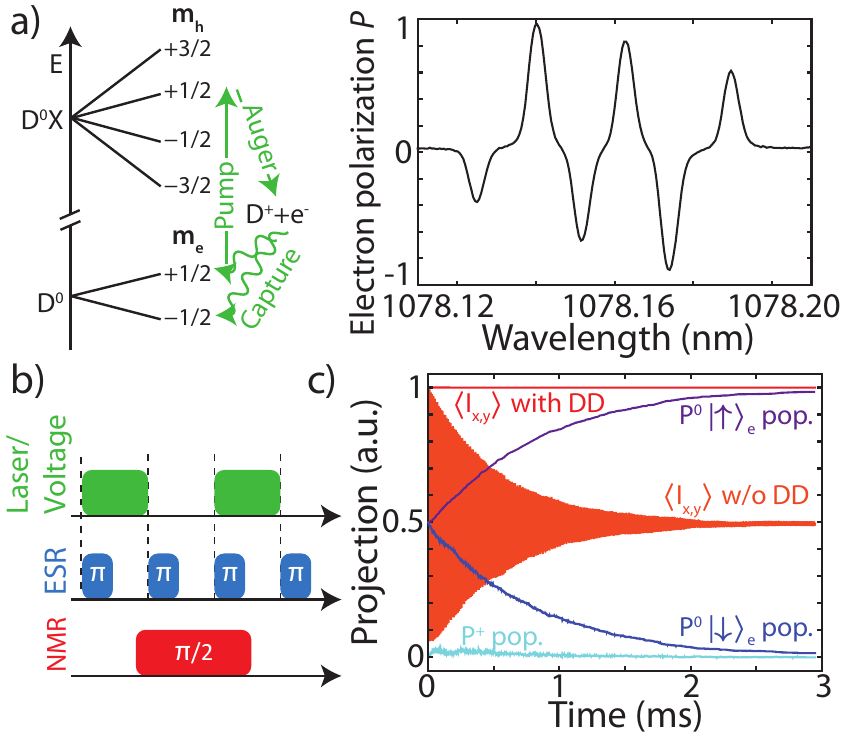}%
	\caption{\textbf{Resetting the donor electron spins whilst preserving nuclear coherence}
	(a)~Donor bound exciton (D$^0$X) energy diagram and measurement of electron spin hyperpolarisation $P=\frac{S_{\mathrm{Echo\ polarized}}}{S_{\mathrm{Echo\ thermal}}} \times \tanh(\frac{h f}{k_BT})$, of donors in \natsi, where $S_{\mathrm{Echo\ polarized}}$ and $S_{\mathrm{Echo\ thermal}}$ are the electron spin echo signal intensities with and without illumination. ESR frequency,  $f=9.7$~GHz, $B$= 349~mT, $T = 4.5$~K. The actual spin polarisation might be somewhat smaller as the enhancement observed could include a contribution for donor nuclear spin polarisation, due to cross-relaxation. 
	(b) Sequence for protecting a weakly coupled \sitwonine\ nuclear spin coherence during donor spin hyperpolarisation by spin-dependent tunneling (voltage pulses, low level = ``read" and high level = ``load'' according to Ref~\citep{Morello2010,Pla2012}) or spin-selective optical ionisation (laser pulses). DD on the donor electron spin (ESR) is synchronised with the laser/voltage pulses in order not to disturb the electron spin polarisation process. DD on the nuclear spin (NMR) is a WAHUHA-like sequence with $\pi/2$ pulses to refocus the dipolar interaction, protecting the nuclear spin coherence from flip-flops when the donor is ionised.
	(c)~Simulation of the sequence in (b) in the case of spin-dependent tunneling, showing the evolution of the donor electron spin and charge states, and the \sitwonine\ nuclear spin coherence with and without DD on the electron spin. Electron spin populations are plotted after every other ESR $\pi$ pulse. Simulation parameters: donor ionisation and capture characteristic times are 295~$\mu$s and 33~$\mu$s, respectively (taken from Ref~\citep{Pla2012}). $\pi$-pulse decoupling rate is 5~MHz for a hyperfine interaction strength of 0.1~MHz. Spin-selective optical ionisation (laser pulses) shows similar evolution but on longer timescales (10--100~ms).
	}
	\label{fig:DDprotection}%
\end{figure}

The long coherence times measured above demonstrate that nuclear spins near the donor could be used as a quantum register, however, applications such as quantum error correction require the ability to repetitively initialise the states of ancilla qubits. Even at low temperatures ($<100$~mK) and high magnetic fields ($> 1$~T), the nuclear spins are in a fairly mixed state in thermal equilibrium, however, the polarisation of the donor electron spin can be transferred to the nuclear spins, following the same methods used in the ENDOR experiments above. Two methods to polarise donor electron spins quickly and on-demand include i) the use of spin-selective donor ionisation through the use of the bound-exciton IR transition (applicable in both ensembles and single spins)~\cite{Yang2006, Steger2011, Lo2014a}; and ii) the measurement of a single donor spin coupled to a single electron transistor (SET)~\cite{Morello2010}. In the first case, laser excitation (at around 1078~nm for \phosphorus) causes only donors of a defined spin orientation to be ionised, which is followed by a subsequent capture of an electron in a random spin state. This can achieve full donor electron spin polarisation on the tens of millisecond timescale (depending on laser power). Although the strain caused by the isotopic variation in natural silicon leads to a broadening of the donor-bound exciton linewidth, the electron spin can still be resolved at modest fields (see \Fig{fig:DDprotection}(a)). In the second case, the timescales are set by tunneling rates between the donor and the SET, which give a measure/reset time of order 1~ms. 

Both of these spin initialisation methods rely on ionisation of the donor, which impacts the coherence of any coupled nuclear spins in two distinct ways. First, while the donor is ionised there is no longer a `frozen core' of protected nuclear spins and so the flip-flops in the nuclear spin bath limit \ttwon\ to the 5~ms timescale~\cite{Dementyev2003}. During such periods, DD sequences similar to WAHUHA can be applied to suppress the dipolar interaction between the spins, as was already demonstrated using NMR in Ref.~\cite{Ladd2005} where the \sitwonine\ nuclear spin coherence was extended up to 20~s. A second issue arises from the inherent uncertainty in the precise timing of the ionisation/neutralisation of the donor, as this imparts a random phase on the nuclear spin related to the strength of its hyperfine coupling to the donor electron.
If the nuclear spin state is an eigenstate, it is rather insensitive to the donor ionization, as evidenced by both optical and electrical ionisation experiments~\cite{Saeedi2013,Pla2013}, however while it is in a superposition state one can expect the random timings of the donor electron removal/re-capture to lead to decoherence. Notably, this decoherence process is also observed in nuclear spins near NV centres in diamond where prolonged measurement of the NV centre can cause it to randomly change its charge state~\cite{Dutt2007}.

One solution is to use nuclear spins whose coupling to the donor electron spin is much weaker than the inverse of the ionisation time uncertainty, but this would require using \sitwonine\ with hyperfine values $\ll 1$~kHz, which in turn have short coherence times and whose conditional operations through the donor electron spin would be slow.
We hence suggest protecting the nuclear spin coherence by applying DD on the electron spin, at times when ionisation/neutralisation of the donor is expected. The hyperfine interaction can thus be effectively turned off on-demand, assuming that the pulses are applied at a repetition rate much faster than the hyperfine coupling strength (see Supplementary Material for derivation). 
Critically, the hyperpolarisation control (in the form of laser or voltage pulses) must be synchronised with the DD pulses in order to work effectively, as illustrated in \Fig{fig:DDprotection}(b). Following this protocol, the electron spin state can be reinitialised while the coherence of (weakly coupled) nuclear spins remains unperturbed (see \Fig{fig:DDprotection}(c)). Finally, this DD method could have further applications, such as protecting the nuclear spins from \tonee\ relaxation of the electron spin (similar to Ref~\citep{Maurer2012}).

Further considerations (see Supplementary Material) for the implementation of a quantum register based on \sitwonine\ weakly coupled to the donor include i) the effect of anisotropy in the hyperfine coupling, and ii) shot-to-shot fluctuations in the state of the nuclear spin bath (manifest as a ESR linewidth of $\approx 8$~MHz~Ref~\cite{Pla2012}). The former could lead to undesired nuclear spin flips as a result of DD applied to the donor electron spin, and can be mitigated by increasing the magnetic field strength. The latter shifts the ESR frequency, however, conditional operations on the electron spin controlled by the nuclear spin can still be performed through the use of frequency comb methods (similar to Mims ENDOR~\cite{Mims1965}).

In conclusion, we have considered the suitability of \sitwonine\ nuclear spins around a donor electron spin as a quantum register, and measured their coherence times to be in the seconds timescale and a function of their hyperfine coupling to the donor. These could be harnessed to perform, for example, a three-qubit QEC protocol using the donor nuclear spin and one strongly coupled \sitwonine\ as ancillae, and one weakly-coupled \sitwonine\ for the data qubit. Combined with recent measurements which show that bismuth donor \emph{electron} spin coherence times can reach a second in natural silicon~\cite{Ma2015}, these results indicate that isotopically enriched \sitwoeight\ may not be a panacea for silicon-based qubits, and the more abundant and easily accessible variant may bring benefits for some applications. Although more technically complex, there may also be merits in incorporating \sitwonine\ in the vicinity of the donor (e.g. through co-implantation), in material which is otherwise isotopically enriched.

We thank C.C.~Lo, S.~Balian, T.~Monteiro, P.~Ross and A.M.~Tyryshkin for valuable discussions and assistance with experiments. This research is supported by the EPSRC through the Materials World Network (EP/I035536/1) and a DTA, as well as by the European Research Council under the European Community's Seventh Framework Programme (FP7/2007-2013) / ERC grant agreement no.\ 279781. J.J.L.M. is supported by the Royal Society.

\bibliography{library}


\end{document}